\documentclass[prb,twocolumn]{revtex4}
\usepackage{amssymb}
\usepackage{graphicx}
\usepackage{amsfonts}
\usepackage{amsmath}

\begin{document}

\title{Electrical switching and interferometry of massive Dirac particles in topological insulators constrictions}

\author{F. Romeo$^{1,2}$ and R. Citro$^{1,2}$}
\affiliation{$^{1}$Dipartimento di Fisica ``E. R. Caianiello'',
Universit{\`a} degli Studi di Salerno,  $^{2}$Institute CNR-SPIN, UO Salerno,  Via Ponte don Melillo,
I-84084 Fisciano (Sa), Italy}
\author{D. Ferraro$^{3,4,5}$ and M. Sassetti$^{3,4}$}
\affiliation{$^{3}$ Dipartimento di Fisica, Universit\`{a} di Genova,Via Dodecaneso 33, 16146, Genova, Italy\\
$^{4}$ CNR-SPIN, Via Dodecaneso 33, 16146, Genova, Italy\\
$^{5}$ INFN, Via Dodecaneso 33, 16146, Genova, Italy.}

%\date{\today}

\begin{abstract}
We investigate the electrical switching of charge and spin transport in a topological insulator nanoconstriction in a four terminal device. The switch of the edge channels is caused by the coupling between edge states which overlap in the constriction and by the tunneling effects at the contacts and therefore can be manipulated by tuning the applied voltages on the split-gate or by geometrical etching.  The switching mechanism can be conveniently studied by electron interferometry involving the measurements of the current in different configurations of the side gates, while the applied bias from the external leads can be tuned to obtain pure charge or pure spin currents (charge- and spin- bias configurations). Relevant signatures of quantum confinement effects, quantum size effects and energy gap are evident in the Fabry-P\'{e}rot physics of the device allowing for a full characterization of the charge and spin currents. The proposed electrical switching behavior offers an efficient tool to manipulate topological edge state transport in a controllable way.
\end{abstract}

%\pacs{73.23.-b}
%73.23.-b electron transport in mesoscopic system
%72.25.Pn spin pump current-driven

%\keywords{topological insulators, edges states, quantum transport of charge and spin}

\maketitle

\section{Introduction}
\label{sec:intro}

The discovery of Topological Insulator systems (TIs), both in three and in two dimensions, has recently attracted enormous attention \cite{Hasan10, Qi11}.  TIs possess an insulating bulk gap and metallic edge or surface states, which can be distinguished from an ordinary band insulator by the existence of $Z_2$ topological invariant\cite{day_phystoday_2008}. This exceptional property leads to quantum spin Hall (QSH) effect which was first proposed for a model graphene system by Kane and Mele\cite{kane_prl_2005}.  As a specific example of two dimensional QSH system, a HgTe/CdTe quantum well (QW) with an inverted band structure has been demonstrated, experimentally and theoretically, to have a single pair of helical edge states in the QSH bar by the appearance of a quantized conductance plateau when the Fermi energy lies in the bulk gap \cite{bernevig_science_2006, Konig07, roth_science_2009}. Quantized transport along the HgTe boundaries can be conveniently explained by an edge channel picture: Two states with opposite spin orientation propagate along opposite device edges in the same direction thus leading to a quantization\cite{Buttiker09} of the conductance of $2e^2/h$. Due to the spatial separation of the spin-states in these systems and to their one-dimensional (1d) nature, system geometry and interference phenomena can be conveniently used for spin selections and HgTe-based topological insulators appear to be promising candidates for spin processing devices. Recent proposals of spin-transistors based on two-dimensional topological insulators rely on the application of a magnetic field at a pn-junction\cite{beenakker_prb_2009}, Aharonov-Bohm and Fabry-P\'{e}rot interferometers \cite{qi_prb_2010,dolcini_prb_2011,noi_prb_2011}, and gating of a single HgTe nanoconstriction \cite{richter_prl_2011, chang_prb_2011}.

In this paper we demonstrate how topological edge states can be electrically switched in an elongated constriction leading to charge and spin transport with high fidelity. The physical mechanisms utilized here involve the coupling between topological edge states (TESs) and the tunneling effects (including spin-flip tunneling) at the extremes of the constriction due to local etching. Both the coupling between the TESs and interference phenomena along the constriction can be controlled by all electrical gating and the present analysis sheds light on how manipulate edge-state transport in TI.\\
In the following we introduce a 1d effective model. Usually the QSH physics in HgTe/CdTe QW was studied by an effective 4-band model that depicts the inversion crossing of electron and hole band\cite{bernevig_science_2006} and most of the works have been based on the numerical solutions within a tight-binding method\cite{qi_prb_2006}, while an analytical solution for the case of a finite strip geometry recently appeared \cite{niu_prl_2008, gong_epl_2011}. The transverse finite size effect is relevant because the edge states on the two sides can couple together to generate a gap in the spectrum even in the clean limit, breaking the edge channels. Since a gap opens in the spectrum, differently from the quantum Hall edge states which do not couple across the width of the strip, the application of electrical gate potentials can be exploited to shift the position of the electrochemical potential within the gap, thus permitting the switching of spin and charge transport. The transport properties of QSH systems in presence of an extended contact have been also considered in the interacting case both to extract information about the intensity of the interaction \cite{Dolcetto12} and to investigate the limit of extremely narrow constrictions \cite{Liu11}.\\
The organization of the paper is the following. In Sec.\ref{sec:ham} we introduce an effective one-dimensional model to describe edge states at the surface of a 2D topological insulator and their coupling along a nanoconstriction. Here we also present the model for the four terminal set-up and its operational configuration. In Sec.\ref{sec:scattering-field} we introduce the scattering field approach and the basic formalism used to calculate spin and charge current in terms of the scattering matrix elements. In Sec.\ref{sec:results} we show the results concerning charge and spin currents, mainly focussing on two specific configurations of the four terminal setup. Finally, we discuss our conclusions in Section \ref{sec:conclusions}.

\section{Model Hamiltonian}
\label{sec:ham}

\begin{figure}[h]
\centering
\includegraphics[scale=0.5]{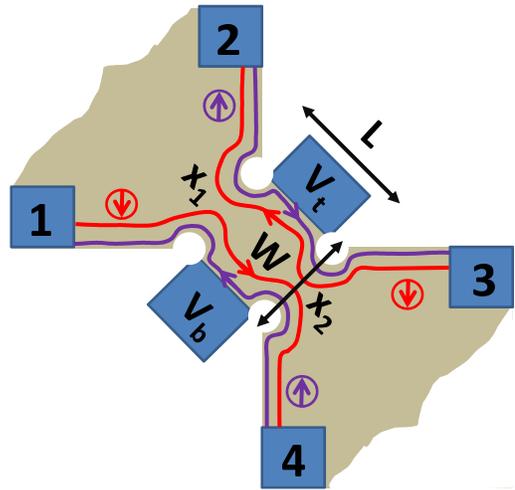}\\
\caption{(Color online) Representation of the edge states flow at the top and bottom boundaries of a TI quantum well. In the middle region, characterized by transverse dimension $W$ and length $L$, a finite overlap among the surface states allows the formation of a gap in the quasi-particle spectrum, while, close to the leads, states with different helicity are uncoupled. The transport properties of the system can be controlled by side gates, i.e. $V_t$ and $V_b$. In particular two gate configurations are analyzed: (i) $V_t=V_b=V_g$; (ii) $V_t=-V_b=V_g$. Two bias configurations are considered: (i) \textit{Charge-bias} defined as $V_1=V_2=V$, $V_3=V_4=0$; (ii) \textit{Spin-bias} defined as $-V_1=V_2=V$, $V_3=V_4=0$.}
\label{fig:device}
\end{figure}

We consider a QSH bar with a nanoconstriction of transverse dimension $W$ and length $L$ formed by a split gate or by a geometrical etching [see Fig.\ref{fig:device}]. Starting from the 4-band model of Refs.[\onlinecite{bernevig_science_2006, Konig07, roth_science_2009}] one can derive an effective Hamiltonian for an infinite strip of constant width $W$. In the region close to the external leads the strip  is characterized by a wide transverse dimension larger than the transverse decay length of the edge state wavefunction. Here one can derive an effective Dirac Hamiltonian  for the edge state $H_{eff,\uparrow/\downarrow}=c\mp \hbar v_{F}\sigma_x k_x$ ($\sigma_x$ being the Pauli matrix) in a single spin sub-block with a velocity $v_{F}$ which corresponds to the Fermi velocity and energy offset $c$ which is in agreement  with the full band structure in the vicinity of the band crossing \cite{richter_prl_2011}.
For decreasing width $W$, i.e. along the constriction, the edge states at opposite boundaries start to overlap, leading to a mass like gap in the particles spectrum whose size is an  exponentially decreasing function of the width $\Gamma \propto \exp(-\lambda W)$, as explained in Appendix \ref{app: appA}, while $\lambda$ is determined by the secular equation for the eigenvalues and depends on the distribution of the wavefunctions in space\cite{niu_prl_2008}. Additionally, one could also take into account the leading order spin-orbit interaction (SOI) due to bulk or structure inversion asymmetry but the overlap of edge states due to both of them is negligibly small and is relevant only close to the avoided band crossing and for constrictions of width of several tens of nanometers ($<100$ nm). We can thus consider the following 1d effective Hamiltonian for a Kramers pairs edge states along the quantum well:
\begin{equation}
\label{eq:hamiltonian}
H=H_0+H_{sp}+H_{sf}+H_c,
\end{equation}
%=====================================================================================fig1
%=========================================================================================
where\cite{note1}:
\begin{eqnarray}
H_0=-i\hbar v_F \sum_{\sigma=\uparrow,\downarrow}\int dx && \lbrack : \psi^\dagger_{R\sigma}(x)\partial_x \psi_{R\sigma}(x): \nonumber \\
&&  -: \psi^\dagger_{L\bar{\sigma}}(x)\partial_x \psi_{L\bar{\sigma}}(x):\rbrack,
\end{eqnarray}
and $\psi_{R(L)\sigma}$ represents the right (left) mover electron annihilation operator with spin $\sigma=\uparrow,\downarrow$, while $:\mathcal{O}:$ stands for the normal ordering of the operator $\mathcal{O}$ with respect to the equilibrium state defined by occupied energy levels below the Fermi sea. In our description, without loss of generality, we assume that spin-$\uparrow$ right movers $(R, \uparrow)$ and spin-$\downarrow$ left movers  $(L, \downarrow)$  flow along the top boundary while the spin-$\downarrow$ right movers $(R, \downarrow)$ and spin-$\uparrow$ left movers $(L, \uparrow)$ flow along the bottom boundary.
Along the nanoconstriction  the edge modes are coupled by confinement effect and the overlap between edge states belonging to different boundaries open a gap in the energy spectrum of the Dirac Fermions.
At the extremes of the constriction ($x=x_1$ and $x=x_2$ ) inter-boundary tunneling events may take place and the only terms which preserve time-reversal symmetry\cite{zhang_ti_2006} can be distinguished in spin-preserving and spin-flipping tunneling described by the following Hamiltonians:
\begin{eqnarray}
\label{eq:backscattering}
H_{sp}=\sum_{\sigma=\uparrow,\downarrow} \int dx &&\lbrack \Gamma_{sp}(x) \psi^\dagger_{R\sigma}(x)\psi_{L\sigma}(x)+\nonumber \\
&&+\Gamma_{sp}(x)^\ast \psi^\dagger_{L\bar{\sigma}}(x)\psi_{R\bar{\sigma}}(x)
\rbrack,
\end{eqnarray}
\begin{eqnarray}
\label{eq:spin-flip}
H_{sf}=\sum_{\alpha=L,R} \int dx &&\xi_\alpha \lbrack \Gamma_{sf}(x) \psi^\dagger_{\alpha\uparrow}(x)\psi_{\alpha\downarrow}(x)+\nonumber \\
&&+\Gamma_{sf}(x)^\ast \psi^\dagger_{\alpha\downarrow}(x)\psi_{\alpha\uparrow}(x)
\rbrack,
\end{eqnarray}
where $\alpha=\{L,R\}$,  $\xi_R=+1,\xi_L=-1$ is the chirality, while $\Gamma_{sf(sp)}(x)$ are the space-dependent tunneling amplitudes:
\begin{equation}
\Gamma_{sf(sp)}(x) = 2\hbar v_F\sum_{i\in 1,2}\gamma_{sf(sp)}\delta(x-x_i).
\end{equation}
Finally the term $H_c$ in (\ref{eq:hamiltonian}) describes the coupling between the edges along the constriction:
\begin{eqnarray}
\label{eq:coupling}
H_{c}=\sum_{\sigma=\uparrow,\downarrow} \int dx && \Gamma(x) \lbrack \psi^\dagger_{R\sigma}(x)\psi_{L\sigma}(x)+\nonumber \\
&&+\psi^\dagger_{L\bar{\sigma}}(x)\psi_{R\bar{\sigma}}(x)\rbrack,
\end{eqnarray}
where $\Gamma(x)=\mathcal{C} (x)\Gamma$ and $\mathcal{C} (x)$ is a step-like function taking value 1 along the nanoconstriction (i.e. $x_1<x<x_2$) and zero elsewhere. Differently from Ref.[\onlinecite{richter_prl_2011}], we assume that the main spin-flipping mechanism in our model is caused by the local modification of the spin-orbit coupling\cite{Vayrynen11}  at $x=x_{1,2}$ governed by $\gamma_{sf}$, while we disregard the spin-orbit interaction eventually present along the constriction $\Gamma_{f}$. The latter assumption is fully justified for not too tight $W$ where $\Gamma_f/\Gamma \ll 1$\cite{richter_prl_2011}.

In the presence of side gates $V_{t}$, $V_{b}$ at top and bottom boundaries of the device, an additional term appears in the Hamiltonian:
\begin{eqnarray}
\label{eq:gates}
H_g=\int_{x_1}^{x_2} dx \lbrack e V_{t} (\rho_{R \uparrow}+\rho_{L\downarrow})+ \nonumber \\
e V_{b}(\rho_{R \downarrow}+\rho_{L\uparrow}) \rbrack,
\end{eqnarray}
where $\rho_{\alpha \sigma}=:\psi^\dagger_{\alpha \sigma}\psi_{\alpha \sigma} :$ denotes the electron density with $\alpha=L,R$ and spin $\sigma$.
The presence of such voltages shifts the edge state momenta in the top and bottom region between $x_1$ and $x_2$, modifying the electron phase in the loop processes induced by the tunneling  and can give rise to electron interference phenomena reminiscent of the Fabry-P\'{e}rot (FP) and Aharonov-Bohm (AB) quantum phases, with respective value given by $\pi \phi_{FP}=e(V_{t}+V_{b})L/\hbar v_F$ and $\pi \phi_{AB}=e(V_{t}-V_{b})L/\hbar v_F$\cite{dolcini_prb_2011}.\\
In the subsequent analysis we consider different bias and gate configurations\cite{chamon_ti_2009} resulting in four possible operational modes of the device. Concerning the bias applied to the four terminals (see Fig. \ref{fig:device}), we define the \textit{charge-bias} configuration (CBC) with  $V_1=V_2=V$, $V_3=V_4=0$, and the \textit{spin-bias} configuration (SBC) corresponding to $-V_1=V_2=V$, $V_3=V_4=0$.
The definition of  `charge' and `spin' configuration originates from the degree of freedom injected through the scattering region, depicted
in Fig.\ref{fig:device}. Indeed in configuration CBC the amount of spin-up and spin-down electrons injected from
terminals 1 and 2 is the same, so that  only the charge degree of freedom is injected, and no net spin.
In contrast, in configuration SBC the lead 1 is negatively biased, determining a depletion of spin-down electrons with
respect to the equilibrium situation, supplying a spin degree of freedom to the arrival leads (i.e. the leads 3 and 4). The same configurations have been discussed in Ref.[\onlinecite{dolcini_prb_2011}] in a model without coupling between the edges. In intermediate situations, i.e. when $|V_1| \ne |V_2|$, both charge and spin degrees of freedom are involved, but for clarity we focus only on CBC and SBS where only {\it pure charge} or {\it pure spin}  current can be generate.
Moreover, two side gates configurations are analyzed: (i) $V_t=V_b=V_g$ ($\phi_{FP}\neq 0 $ and $ \phi_{AB}=0 $); (ii) $V_t=-V_b=V_g$ ($\phi_{FP}=0 $ and $ \phi_{AB} \neq 0 $), which permit us to analyze electron interferometric phenomena.

\section{Scattering fields approach}
\label{sec:scattering-field}

We now formulate a scattering field theory \textit{\`{a} la} B\"{u}ttiker\cite{buttiker_92} able to describe coherent spin and charge transport in the system shown in Fig.\ref{fig:device}.\\
The charge or spin current operators $\hat{J}_{c/s}$ in first quantization are written as follows:
\begin{eqnarray}
\hat{J}_{c} &=& v_F e\hat{\tau}_z\otimes \mathbb{I}_{2 \times 2}\\\nonumber
\hat{J}_{s} &=& v_F \frac{\hbar}{2}\hat{\tau}_z\otimes \hat{\sigma}_{z},
\end{eqnarray}
where $\hat{\sigma}_{z}(\hat{\tau}_{z})$ stands for the Pauli matrix, $ \mathbb{I}_{2 \times 2}$ for the identity matrix acting on the Hilbert space given by the tensor product $| \alpha\rangle\otimes|\sigma\rangle$ ($\alpha \in \{R,L\}$, $\sigma \in \{\uparrow,\downarrow\}$).
To build a scattering field theory, one first defines the scattering field corresponding to each terminal $i=1,\ldots,4$ in terms of the incoming ($\hat{a}_{\alpha\sigma}(E)$) and outgoing ($\hat{b}_{\alpha\sigma}(E)$) electron operators, according to:
\begin{eqnarray}
\hat{\Psi}_1(x,t) &=& \int dE \frac{e^{-iEt/\hbar}}{\sqrt{h v_F}}\Bigl[\hat{a}_{R\downarrow}(E;x)+\hat{b}_{L\uparrow}(E;x)\Bigl]\\\nonumber
\hat{\Psi}_2(x,t) &=& \int dE \frac{e^{-iEt/\hbar}}{\sqrt{h v_F}}\Bigl[\hat{a}_{R\uparrow}(E;x)+\hat{b}_{L\downarrow}(E;x)\Bigl]\\\nonumber
\hat{\Psi}_3(x,t) &=& \int dE \frac{e^{-iEt/\hbar}}{\sqrt{h v_F}}\Bigl[\hat{a}_{L\downarrow}(E;x)+\hat{b}_{R\uparrow}(E;x)\Bigl]\\\nonumber
\hat{\Psi}_4(x,t) &=& \int dE \frac{e^{-iEt/\hbar}}{\sqrt{h v_F}}\Bigl[\hat{a}_{L\uparrow}(E;x)+\hat{b}_{R\downarrow}(E;x)\Bigl],
\end{eqnarray}
where $\hat{a}_{\alpha\sigma}(E;x)=\hat{a}_{\alpha\sigma}(E)| \alpha\rangle\otimes|\sigma\rangle \exp(i\eta_{\alpha} k_E x)$ with $\eta_{R}=-\eta_{L}=1$ and the wavevector $k_E=E/(\hbar v_F)$ (and similarly for $\hat{b}$) .\\
The second-quantized current operators in the terminal $i$ is defined by $\hat{\mathbf{J}}^{(i)}_{c/s}=\hat{\Psi}_i^{\dag}\hat{J}_{c/s}\hat{\Psi}_i$ and is explicitly given by ($\mu \in \{c,s\}$):
\begin{equation}
\label{eq:current_op}
\hat{\mathbf{J}}^{(i)}_{\mu}=\epsilon_i g_{\mu}\Bigl[ (\xi_{\mu})^{i+1}\hat{a}^{\dag}_{i}\hat{a}_{i}+(\xi_{\mu})^{i}\hat{b}^{\dag}_{i}\hat{b}_{i}\Bigl],
\end{equation}
where $g_{c}=|e|/h$, $g_{s}=1/(4\pi)$, $\xi_{c/s}=\mp 1$ and $\epsilon_{1,4}=-1=-\epsilon_{2,3}$. In writing Eq.~(\ref{eq:current_op}) we made use of the Fourier transform $\hat{a}_i(t)=\int dE \hat{a}_{i}(E)\exp[-iEt/\hbar]$, while the following correspondence has been made:
$[b_1, b_2, b_3, b_4]^t=[b_{L\uparrow}, b_{L\downarrow}, b_{R\uparrow}, b_{R\downarrow}]^t$, $[a_1, a_2, a_3, a_4]^t=[a_{R\downarrow}, a_{R\uparrow}, a_{L\downarrow}, a_{L\uparrow}]^t$.\\
The expectation value $\langle \hat{\mathbf{J}}^{(i)}_{\mu}\rangle$ can be computed making use of the scattering relation $b_j=\sum_{i}S_{ji}a_i$ and from quantum statical average $\langle \hat{a}^{\dag}_{j}(E) \hat{a}_{i}(E')\rangle=\delta_{ij}\delta(E-E')f_i(E)$, being $f_i(E)$ the Fermi-Dirac distribution with electrochemical potential $\mu_j=\tilde{\mu}+eV_j$. After direct computation we get:
\begin{equation}
\label{eq:current_exp}
\langle\hat{\mathbf{J}}^{(i)}_{\mu}\rangle=\epsilon_i g_{\mu}\int dE \sum_{j}\Bigl[\delta_{ij}(\xi_{\mu})^{i+1}+(\xi_{\mu})^i|S_{ij}(E)|^2\Bigl]f_{j}(E).
\end{equation}
In the linear response regime $f_j(E)$ can be expanded around the equilibrium energy  $\varepsilon=E-\tilde{\mu}$ and for small bias one can express the charge/spin current in terms of a generalized conductance tensor  $\langle\hat{\mathbf{J}}^{(i)}_{\mu}\rangle=\sum_{j}G^{\mu}_{ij}V_j$, whose elements are given by:
\begin{equation}
\label{eq:conductance_tensor}
G^{\mu}_{ij}=\epsilon_i g_{\mu}|e|\int d \varepsilon\Bigl[\delta_{ij}(\xi_{\mu})^{i+1}+(\xi_{\mu})^i|S_{ij}(\varepsilon)|^2\Bigl]\Bigl(\partial_{\varepsilon} f(\varepsilon)\Bigl)_{eq}.
\end{equation}
Eq.~(\ref{eq:conductance_tensor}) provides a description of the linear response theory of the system in terms of the scattering matrix elements. Since we are interested in the quantum regime, we shall limit our analysis to the zero temperature case.

\subsection{Boundary conditions and scattering matrix}
The scattering matrix $S_{ij}$ is a four by four unitary matrix whose diagonal entries vanish by helicity and time-reversal symmetry while all the other entries
can be explicitly determined as a function of the tunneling amplitudes $\gamma_{sp},\gamma_{sf}$ by imposing the proper boundary conditions (BCs) on the wave functions\cite{dolcini_prb_2011, noi_prb_2011}. Since the wavefunctions are continuous in the regions $x<x_1$,  $x_1<x<x_2$ and $x>x_2$, we only have to impose the matching conditions where Dirac delta potentials are present, i.e. at $x=x_{1/2}$. By using the equation of motion of the quantum fields (see Appendix \ref{app: appA}) and explicitly taking into account the properties of the Dirac delta potential under integration, one obtains the following  matching conditions: \begin{equation}
\label{eq:BCs}
\mathcal{M}\Psi(x_i+0^{+})=\mathcal{M}^{\ast}\Psi(x_i-0^{+})
\end{equation}
($i \in \{1,2\}$)  where the matrix $\mathcal{M}$ is given by:

\begin{equation}
\label{eq:match}
\mathcal{M}=\left(
              \begin{array}{cccc}
                1 & i\gamma_{sf} & i \gamma_{sp} & 0 \\
                i \gamma_{sf} & 1 & 0 & i \gamma_{sp} \\
                -i\gamma_{sp} & 0 & 1 & i \gamma_{sf} \\
                0 & -i \gamma_{sp} & i\gamma_{sf} & 1 \\
              \end{array}
            \right).
\end{equation}
In the limit of vanishing $\gamma_{sf}$ and $\gamma_{sp}$, $\mathcal{M}$ becomes the identity matrix $\mathbb{I}_{4 \times 4}$ and thus the BCs simply require the continuity of wavefunctions in $x=x_i$. The BCs in (\ref{eq:BCs})  provide 8 equations from which the scattering matrix elements can be numerically determined. Once the scattering matrix is known, the charge and spin currents can be computed by using Eq.~(\ref{eq:current_exp}).

\section{Results}
\label{sec:results}

In the following analysis we study the charge and spin currents induced through the system as the effect of the applied bias $V_i$, $i \in \{1,...,4\}$. We will work in dimensionless units. In particular, the charge (spin) currents are  expressed in unit of $V e^2/h$ [$eV/(4\pi)$], the energy is measured in unit of $ \Xi=0.65$ meV which corresponds to half gap in the coupling region for the parameters here used, while the distance $L$ is rendered dimensionless by the substitution $ L \rightarrow \Xi L /(\hbar v_F) \equiv d$ (notice that $L\approx 0.49 \ \mu$m $\times d$, for $v_F \approx 0.48 \cdot 10^{6}$ m/s). Finally, we take the coupling $\Gamma=2$ corresponding to $1.3$ meV, being it an appropriate value for a nanoconstriction of $W=100$ nm (see Appendix \ref{app: appA}).
The currents are measured in the terminals 3 and 4, which are assumed to be grounded, i.e. $f_3 = f_4 = f_{eq}$, with $f_{eq}$
the Fermi distribution at equilibrium, while the $J_{c/s}$ \textit{vs} $d$ curves are shown on a very large range (i.e. $d \in [0.2, 6]$) to better identify the characteristic oscillation scales.

\subsection{Gates configuration 1: $V_t=V_b=V_g$ and bias configuration CBC}

In this configuration the momentum of the edge states along the nanoconstriction is shifted by the Fabry-P\'{e}rot phase and
the energy spectrum, which can be obtained by a straightforward diagonalization of the Hamiltonian $H_{0}+H_{c}+H_{g}$, is characterized by two branches:
\begin{equation}
\label{eq:spectrum1}
E_{\pm}=eV_g \pm \sqrt{(\hbar v_F k)^2+\Gamma^2},
\end{equation}
originating from the coupling $\Gamma$ of left- and right-movers in  each spin channel.
When the Fermi energy is located within the gap, for sufficient long $L$, the particles transport is strongly suppressed,
while applying the gate $V_g$ the edge channels transport can be switched.
In the following we consider the particles transport through the \textit{conduction band}, i.e. the one specified by
the $"+"$ sign in Eq.~(\ref{eq:spectrum1}).\\
%----------------------------------------------------------fig2
\begin{figure}[h]
\centering
\includegraphics[clip,scale=0.65]{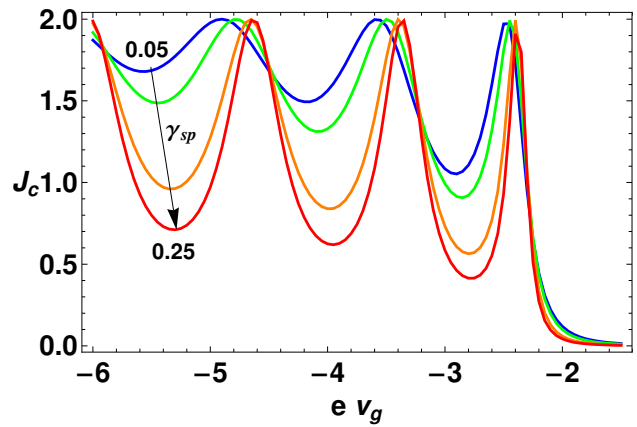}\\
\caption{(Color online) Charge currents $J_c$ vs $eV_g$  in gate configuration 1 computed
by fixing the model parameters as follows: $d=2$, $\Gamma=2$, $\gamma_{sf}=0.1$.
The parameter $\gamma_{sp}$ varies from top to bottom  $\gamma_{sp} \in \{0.05, 0.1, 0.2, 0.25\}$. The above curves, for a fixed value of $\gamma_{sp}$, do not depend on the value of $\gamma_{sf}$.}
\label{fig:fig2}
\end{figure}
%----------------------------------------------------------
In Fig.\ref{fig:fig2} we present the results for the CBC where a pure charge current $J_c$ is generated. By fixing the system parameters as $d=2$, $\Gamma=2$, $\gamma_{sf}=0.1$, we plotted a set of curves of $J_c$ \textit{vs} $eV_g$ at varying $\gamma_{sp}$ in the set $\{0.05, 0.1, 0.2, 0.25\}$. When $eV_g=0$ the conductance is strongly suppressed, while for a finite gate potential a non-vanishing current can be generated. The $J_c$ \textit{vs} $eV_g$ has an oscillatory behavior associated to the Fabry-P\'{e}rot-like resonance in the constriction. The maxima of $J_c$ are determined by the constructive interference condition along the nanoconstriction:
\begin{equation}
\label{eq:interference_con}
2k d =2 \pi n-2\phi_s,
\end{equation}
where $\phi_s$ is a scattering phase not known a priori and which depends on the transparency of the barriers, i.e. on the tunneling amplitude $\gamma_{sp}, \gamma_{sf}$ at $x_{1,2}$, while the particles momentum $k$ is determined by Eq.~(\ref{eq:spectrum1}). The values of the bias maximizing $J_c$ are instead given by
$(eV_g)_n=-\sqrt{\Gamma^2+[(\pi n-\phi_s)/d]^2}$ in correspondence of the Fermi level $E_F=0$. The dependence of $J_c$ on the scattering phase $\phi_s$ is evident in the curves of Fig.\ref{fig:fig2} where $\gamma_{sp}$ is varied from lower to higher values.
In  particular, a linear shift of the interference maxima (minima) accompanied by an higher amplitude modulation (at high values of $\gamma_{sp}$)
is observed. The shifting of maxima of $J_c$ at varying $\gamma_{sp}$ is not observed in the case of Dirac Fermions with linear dispersion relation
($E\propto k$) and thus the observed $\phi_s$-dependent shift is a peculiar feature of gapped Dirac particles. In fact, in this case the non-linear energy dispersion leads to scattering properties analogous to those of massive Schr\"{o}dinger particles.
%----------------------------------------------------------fig3
\begin{figure}[h]
\centering
\includegraphics[clip,scale=0.65]{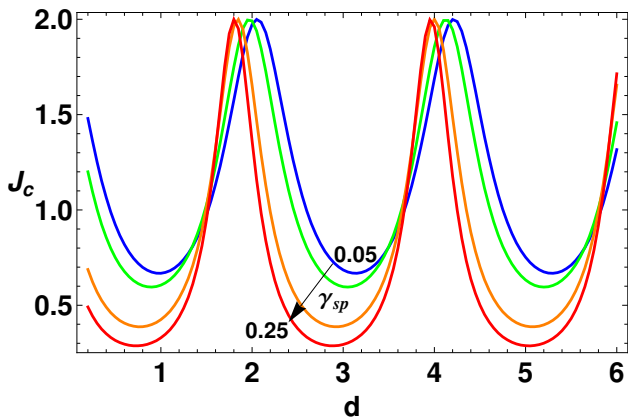}\\
\caption{(Color online) Charge currents $J_c$ vs $d$ in gate configuration 1
computed by fixing the model parameters as follows: $eV_g=-2.45$, $\Gamma=2$, $\gamma_{sf}=0.1$.
The parameter $\gamma_{sp}$ varies from top to bottom in the range $\gamma_{sp} \in \{0.05, 0.1, 0.2, 0.25\}$. }
\label{fig:fig3}
\end{figure}
%----------------------------------------------------------
An experimental study of the oscillating behavior of the $J_c$ \textit{vs} $eV_g$ could provides relevant information
on the coupling energy $\Gamma$ and on the scattering phase $\phi_s$.\\
In Fig. \ref{fig:fig3} we study the charge current as a function of the dimensionless length $d$ of the constriction by fixing the remaining parameters as follows: $eV_g=-2.45$, $\Gamma=2$, $\gamma_{sf}=0.1$, while $\gamma_{sp}$ takes the values $\{0.05, 0.1, 0.2, 0.25\}$. Apart from the shifting of the interference maxima, a general periodic behavior is observed. The space period $\tau_d$ depends on the Fermi energy $E_F$, on the applied voltage $V_g$ and on the coupling $\Gamma$ and can easily be determined by (\ref{eq:spectrum1}) and (\ref{eq:interference_con}):
\begin{equation}
\label{eq:tau_d}
\tau_d=\frac{\pi}{\sqrt{(E_F-eV_g)^2-\Gamma^2}}.
\end{equation}
Let us note that $\tau_d$ does not depend on $\gamma_{sp}$ nor on the scattering phase $\phi_s$  which is instead involved in the interference condition (\ref{eq:interference_con}), providing
a shift of the maxima of $J_c$. Furthermore, the expression of $\tau_d$ can be experimentally
used to determine the average value of the coupling $\Gamma$ when more devices
with different channel length $d$ are at disposal.

\subsection{Gates configuration 1: $V_t=V_b=V_g$ and bias configuration SBC}

The device under consideration (Fig.\ref{fig:device}) can also work
in a different bias configuration allowing for the generation of pure spin current (SBC).
In the following we present results in this configuration.
%----------------------------------------------------------fig4
\begin{figure}[h]
\centering
\includegraphics[clip,scale=0.65]{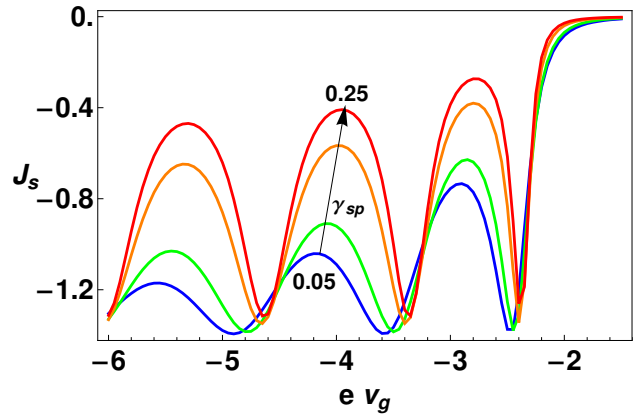}\\
\caption{(Color online) Spin currents $J_s$ vs $eV_g$ in gate configuration 1 computed by fixing the model parameters as follows: $d=2$, $\Gamma=2$, $\gamma_{sf}=0.1$. The parameter $\gamma_{sp}$ varies from bottom to top curve as $\gamma_{sp} \in \{0.05, 0.1, 0.2, 0.25\}$.}
\label{fig:fig4}
\end{figure}
%----------------------------------------------------------
In Fig.\ref{fig:fig4} we show the spin current $J_s$ as a function of $eV_g$ by setting the model parameters
as: $d=2$, $\Gamma=2$, $\gamma_{sf}=0.1$, while the different curves correspond to
$\gamma_{sp} \in \{0.05, 0.1, 0.2, 0.25\}$. The oscillatory behavior as a function of $V_g$ and the periodicity of the curves follows strictly
the one described for the CBC by Eq.(\ref{eq:interference_con}). However, differently from that case a lowering of the maxima
of the spin currents is observed by decreasing $\gamma_{sp}$. This effect is also evident
in Fig.\ref{fig:fig5}, where $J_s$ \textit{vs} $d$ curves are shown by fixing $eV_g=-2.45$,
while maintaining the other parameters as given in Fig.\ref{fig:fig4}.
The lowering of the maxima as a function of $\gamma_{sp}$ can be understood by observing that the expectation value of
the spin current is given by $v_F (\hbar/2)\sqrt{1-\left(\Gamma/E\right)^2}$ which depends
on the scattering phase $\phi_s$ through the energy evaluated at $eV_g$.
%----------------------------------------------------------fig5
\begin{figure}[h]
\centering
\includegraphics[clip,scale=0.65]{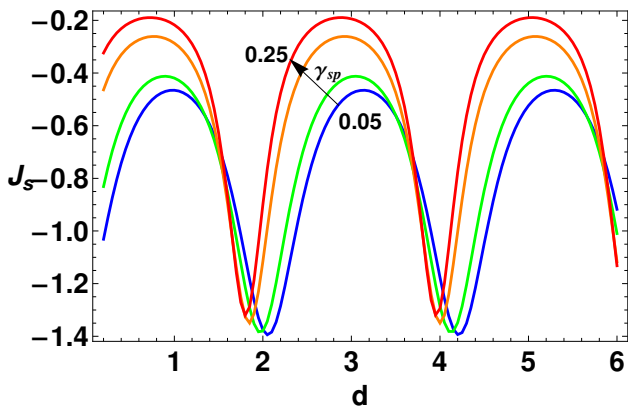}\\
\caption{(Color online) Spin currents $J_s$ vs $d$ in gate configuration 1 computed by fixing the model parameters as follows:
$eV_g=-2.45$, $\Gamma=2$, $\gamma_{sf}=0.1$. The parameter $\gamma_{sp}$ varies from bottom to top curve as $\gamma_{sp} \in \{0.05, 0.1, 0.2, 0.25\}$.}
\label{fig:fig5}
\end{figure}
%----------------------------------------------------------
Differently from the charge current analyzed in CBC, the spin current $J_s$ depends on the value of  $\gamma_{sf}$, the spin-flipping tunneling amplitude. This dependence is evident in Fig.\ref{fig:fig6}, where the spin current $J_s$ is analyzed as a function of the constriction length $d$ by fixing the model parameters as $eV_g=-2.45$, $\Gamma=2$, $\gamma_{sp}=0.1$, while the different curves are computed for $\gamma_{sf} \in \{0.05, 0.1, 0.2, 0.25\}$. As already seen in Fig.\ref{fig:fig5}, an oscillating behavior of the $J_s$ \textit{vs} $d$ curves is observed, the period of the oscillation being described by Eq.~(\ref{eq:tau_d}). However, an important feature in the curves in Fig.\ref{fig:fig6} is that by increasing $\gamma_{sf}$ above a threshold value ($\gamma_{sf}\approx 0.2$) the spin current changes sign (see APPENDIX \ref{app: appB}). Since $\gamma_{sf}$ can be locally controlled by geometrical etching, the change of sign of the spin current can be implemented as a switch for spintronics purposes or to characterize the constriction. The above method is expected to be quite robust against decoherence  phenomena which can reduce the Fabry-P\'{e}rot oscillations amplitude while unaffecting the mean value of the current.\\

%----------------------------------------------------------fig6
\begin{figure}[h]
\centering
\includegraphics[clip,scale=0.65]{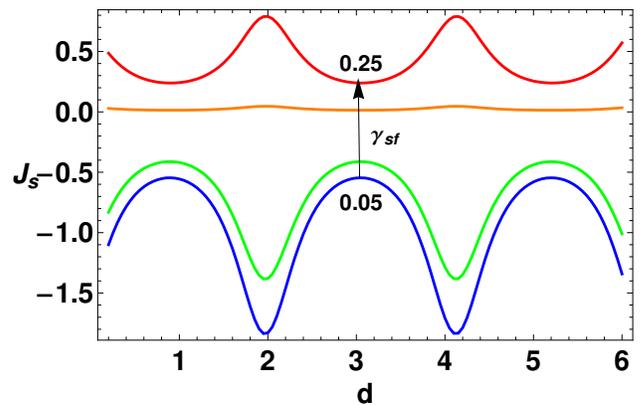}\\
\caption{(Color online) Spin currents $J_s$ vs $d$ in gate configuration 1 computed by fixing the model parameters as follows: $eV_g=-2.45$, $\Gamma=2$, $\gamma_{sp}=0.1$. The parameter $\gamma_{sf}$ varies from bottom to top curve as $\gamma_{sf} \in \{0.05, 0.1, 0.2, 0.25\}$. See also APPENDIX \ref{app: appB}.}
\label{fig:fig6}
\end{figure}
\begin{figure}[h]
\centering
\includegraphics[clip,scale=0.65]{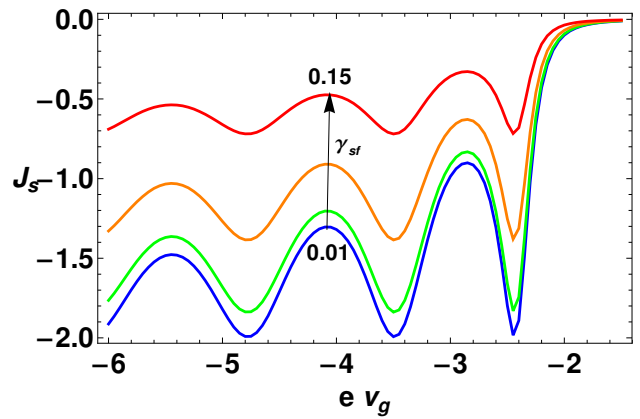}\\
\caption{(Color online) Spin currents $J_s$ vs $eV_g$ in gate configuration 1 computed by fixing the model parameters as follows:
$d=2$, $\Gamma=2$, $\gamma_{sp}=0.1$. The parameter $\gamma_{sf}$ varies from bottom to top curve as $\gamma_{sf} \in \{0.01, 0.05, 0.1, 0.15\}$.}
\label{fig:fig7}
\end{figure}

%----------------------------------------------------------
In Fig.\ref{fig:fig7} we analyze the Fabry-P\'{e}rot oscillations of the spin current $J_s$ as a function of $eV_g$ by fixing the model parameters as
$d=2$, $\Gamma=2$, $\gamma_{sp}=0.1$, the parameter $\gamma_{sf}$ being fixed from bottom to top curve as $\gamma_{sf} \in \{0.01, 0.05, 0.1, 0.15\}$.
As evident from the figure, a decreasing of $\gamma_{sf}$ produces an increasing of the spin polarized current
generated in the system, the maximum value being related to the number $\mathcal{N}$ of  modes involved in the transport (i.e. $\mathcal{N}=2$).\\
%----------------------------------------------------------fig7
%----------------------------------------------------------
As a final remark, we observe explicitly that the charge current in Fig.\ref{fig:fig2} (CBC) and the spin current  in Fig.\ref{fig:fig7} (SBC) can be controlled by using $V_g$. Indeed, starting from $eV_g=-2$ (\textit{off-state}), a sudden activation of the transport along the edge can be induced by tuning $eV_g$ to the value $-2.4$ (\textit{on-state}).

\subsection{Gates configuration 2: $V_t=-V_b=V_g$ and bias configuration CBC}
In this configuration the side gates energy $eV_g$ removes the spin degeneracy leading to the following energy spectrum:
\begin{equation}
\label{eq:spectrum2}
E_{\beta\eta}=\beta \sqrt{(\hbar v_F k+\eta eV_g)^2+\Gamma^2},
\end{equation}
where $\beta=\pm 1$ indicates the conduction ($"+"$) and the valence ($"-"$) band, while $\eta=\pm 1$
describes the momentum shift of particles. The spectrum (\ref{eq:spectrum2}) looks very similar to the one originated by the presence of spin-orbit coupling along the constriction. In particular, the eigenstate corresponding to $\eta=1$ acquires
the phase factor $\exp[-ieV_g/(\hbar v_F)]$ compared to the case with $V_g=0$, while that for
$\eta=-1$ the factor $\exp[ieV_g/(\hbar v_F)]$.
 In the following discussion, we fix the Fermi energy
 $E_F=2.45$ and $\Gamma=2$ to obtain a non-vanishing particles transport through the \textit{conduction band} ($\beta=+1$). The value of $E_F$ is a tunable quantity  \cite{Konig07, roth_science_2009} and can be controlled by using an additional back-gate acting below the whole nanostructure,
 the more interesting regime being the one with $E_F$ just above the gap.
%----------------------------------------------------------fig8
\begin{figure}[h]
\centering
\includegraphics[clip,scale=0.65]{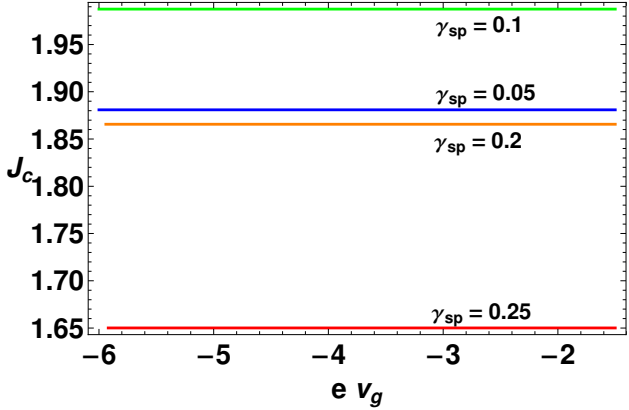}\\
\caption{(Color online) Charge currents $J_c$ \textit{vs} $eV_g$ in gate configuration 2 computed by fixing the model parameters as follows: $d=2$, $\Gamma=2$, $\gamma_{sf}=0.1$, $E_F=2.45$. The parameter $\gamma_{sp}$ varies as $\gamma_{sp} \in \{0.05, 0.1, 0.2, 0.25\}$. The above curves, for a fixed value of $\gamma_{sp}$, do not depend on the value of $\gamma_{sf}$.}
\label{fig:fig8}
\end{figure}
%----------------------------------------------------------
In Fig.\ref{fig:fig8} we study the charge current $J_c$ as a function of the applied gate $eV_g$ by fixing the model parameters as follows:  $d=2$, $\Gamma=2$, $\gamma_{sf}=0.1$, $E_F=2.45$, while $\gamma_{sp} \in \{0.05, 0.1, 0.2, 0.25\}$. A distinctive feature is that $V_g$ does not induce  a charge current modulation as the one observed in Fig.\ref{fig:fig2}. This phenomenon can be explained observing  that the interference effects responsible for a charge current modulation are related to the Fabry-P\'{e}rot phase $\phi_{FP}\propto V_t+V_b$ which is zero in this configuration. The analysis of $J_c$ as a function of the constriction length $d$ is shown in Fig.\ref{fig:fig9} setting the model parameters as follows:  $eV_g=-2.4$, $\Gamma=2$, $\gamma_{sf}=0.1$, $E_F=2.45$, $\gamma_{sp} \in \{0.05, 0.1, 0.2, 0.25\}$. The $J_c$ \textit{vs} $d$ curves show an oscillating behavior whose space period can be deduced by the values of $k$ obtained
from  Eq.~(\ref{eq:spectrum2}) fixing the energy at $E_F$. In this way we obtain two different space periods:
\begin{equation}
\tau^{(\pm)}_d=\frac{\pi}{\sqrt{E_F^2-\Gamma^2}\pm eV_g}
\end{equation}
which define two characteristic frequencies  $\omega_{\pm}=2\pi/\tau^{(\pm)}_d$ that determine the harmonic content of the charge and spin current curves. In particular, as in the familiar case of superposition of waves with different frequencies, one expects to see an oscillating function with frequencies $\Omega_{1/2}=(\omega_+\pm \omega_-)/2$ and whose corresponding space periods are:
\begin{eqnarray}
\label{eq:tau_d_splitgate}
\tau_d^{(1)} &=& \frac{\pi}{e V_g}\\\nonumber
\tau_d^{(2)} &=& \frac{2 \pi}{\sqrt{E_F^2-\Gamma^2}}.
\end{eqnarray}
%----------------------------------------------------------fig9
\begin{figure}[h]
\centering
\includegraphics[clip,scale=0.65]{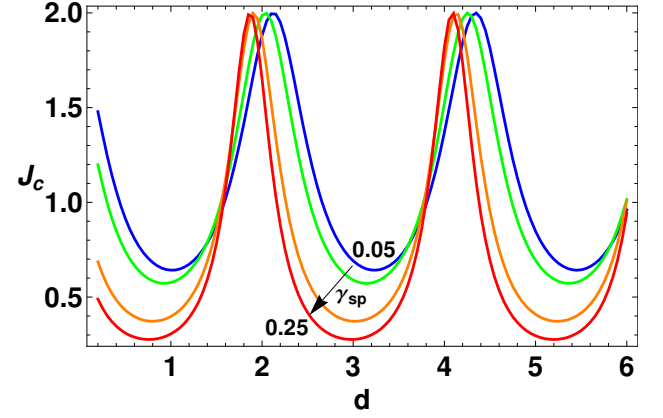}\\
\caption{(Color online) Charge currents $J_c$ \textit{vs} $d$ in gate configuration 2 computed by fixing the model parameters as
follows: $eV_g=-2.4$, $\Gamma=2$, $\gamma_{sf}=0.1$, $E_F=2.45$. The parameter $\gamma_{sp}$ varies as $\gamma_{sp} \in \{0.05, 0.1, 0.2, 0.25\}$.}
\label{fig:fig9}
\end{figure}
%--------------------------------------------------------------
In Fig.\ref{fig:fig9} the period $\tau_d^{(2)}/2 \approx 2.22$ is soon recognized, while the period $\tau_d^{(1)}\approx 1.31 $ is not observable being near equal to a subharmonics of $\tau_d^{(2)}/2$.

\subsection{Gates configuration 2: $V_t=-V_b=V_g$ and bias configuration SBC}

\begin{figure}[h]
\centering
\includegraphics[clip,scale=0.65]{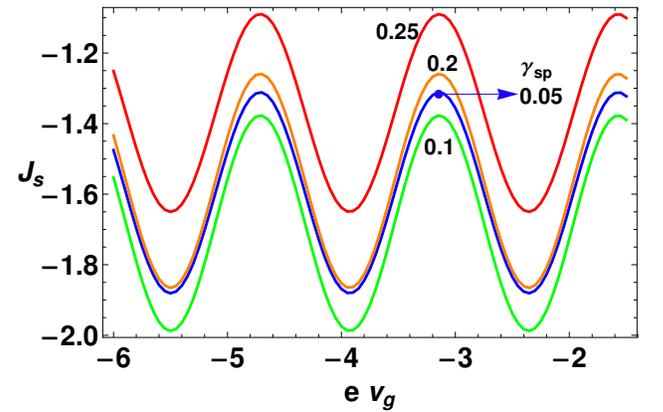}\\
\caption{(Color online) Spin currents $J_s$ \textit{vs} $eV_g$ in gate configuration 2 computed by fixing the model parameters as follows: $d=2$, $\Gamma=2$, $\gamma_{sf}=0.1$, $E_F=2.45$. The parameter $\gamma_{sp}$ varies as $\gamma_{sp} \in \{0.05, 0.1, 0.2, 0.25\}$.}
\label{fig:fig10}
\end{figure}

\begin{figure}[h]
\centering
\includegraphics[clip,scale=0.65]{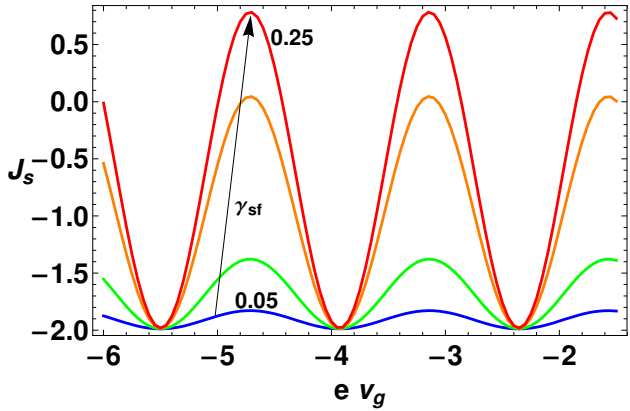}\\
\caption{(Color online) Spin currents $J_s$ \textit{vs} $eV_g$ in gate configuration 2 computed by fixing the model parameters as follows: $d=2$, $\Gamma=2$, $\gamma_{sp}=0.1$, $E_F=2.45$. The parameter $\gamma_{sf}$ varies as $\gamma_{sf} \in \{0.05, 0.1, 0.2, 0.25\}$. }
\label{fig:fig11}
\end{figure}

\begin{figure}[h]
\centering
\includegraphics[clip,scale=0.65]{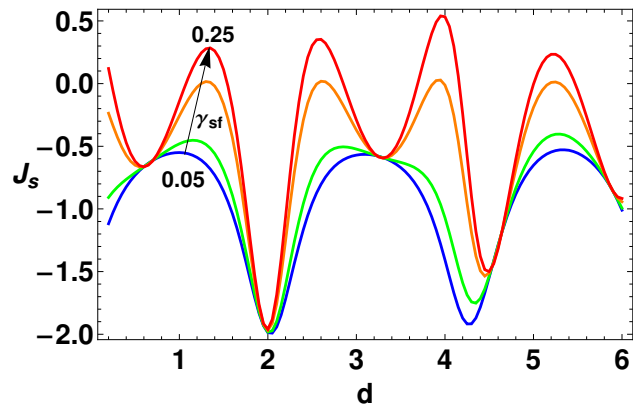}\\
\caption{(Color online) Spin currents $J_s$ \textit{vs} $d$ in gate configuration 2 computed by fixing the model parameters as follows: $eV_g=-2.4$, $\Gamma=2$, $\gamma_{sp}=0.1$, $E_F=2.45$. The parameter $\gamma_{sf}$ varies as $\gamma_{sf} \in \{0.05, 0.1, 0.2, 0.25\}$.}
\label{fig:fig12}
\end{figure}

When the system is driven in the SBC, pure spin currents are generated. In the following we focus on this configuration.\\
In Fig.\ref{fig:fig10} we present the spin current $J_s$ as a function of $eV_g$ by setting the model
parameters as follows:  $d=2$, $\Gamma=2$, $\gamma_{sf}=0.1$, $E_F=2.45$, $\gamma_{sp} \in \{0.05, 0.1, 0.2, 0.25\}$.
The figure shows a gate-induced spin current modulation whose voltage period,
independent from $\gamma_{sp/sf}$, is given by $\tau_v=\pi/d$.
%----------------------------------------------------------fig10
%----------------------------------------------------------
The same period is found in Fig.\ref{fig:fig11} where
 we analyze the spin current $J_s$ as a function of $eV_g$ fixing the model parameters as follows:  $d=2$, $\Gamma=2$, $\gamma_{sp}=0.1$, $E_F=2.45$, $\gamma_{sf} \in \{0.05, 0.1, 0.2, 0.25\}$. In this case a peculiar form of spin-switch occurs. Indeed, at increasing values of $\gamma_{sf}$ the curves show a change of sign
 of the spin current when the gate is tuned close to odd integer values.
%----------------------------------------------------------fig11
%----------------------------------------------------------
However, the possibility to obtain a full sign reversal is determined by the constriction properties
(i.e. $\gamma_{sf}$). To fully characterize the spin-switching phenomenon, in Fig.\ref{fig:fig12} we report the spin current as a function of the constriction length $d$ by fixing the model parameters as follows:  $eV_g=-2.4$, $\Gamma=2$, $\gamma_{sp}=0.1$, $E_F=2.45$, $\gamma_{sf} \in \{0.05, 0.1, 0.2, 0.25\}$. The figure shows a complicated
beating-like pattern whose periods are determined by the ones in Eq.~(\ref{eq:tau_d_splitgate}).
The interesting aspect in Fig.\ref{fig:fig12} is that, in the parameters range investigated,
there exist  special values of $d$ where the corresponding value of $J_s$ is independent on $\gamma_{sf}$.
This property implies
that devices having a value of $d$ close to these special points
do not manifest the spin switching phenomenon.
%----------------------------------------------------------fig12
%----------------------------------------------------------

\section{Conclusions}
\label{sec:conclusions}

We investigated the electrical switching of charge and spin transport in a topological insulator nanoconstriction in a four terminal device by means of a scattering field theory. The spin and charge switch of the edge channels is caused by the coupling between edges states which overlap in the constriction and by the tunneling effects at the contacts of the quantum spin Hall bar and therefore can be manipulated by tuning external applied gate voltages and by geometrical etching. We showed that the switching mechanism can be conveniently studied by electron interferometry involving the measurements of charge and spin currents in specific bias and gate configurations (the switching probability is analyzed in the Appendix \ref{app: appB}).
The device can operate in two gate configurations defined as: (i) $V_t=V_b=V_g$; (ii) $V_t=-V_b=V_g$, with side gates $V_t$ and $V_b$. Furthermore, operating with the external dc bias $V_i$, the device can work in the following configurations: (i) \textit{Charge-bias} (CBC) defined as $V_1=V_2=V$, $V_3=V_4=0$; (ii) \textit{Spin-bias} (SBC) defined as $-V_1=V_2=V$, $V_3=V_4=0$. The CBC and the SBC produce \textit{pure charge} and \textit{pure spin} current, respectively.\\
Concerning the gate configuration (i), we showed that in CBC the Fabry-P\'{e}rot interference maxima of $J_c$ \textit{vs} $eV_g$ depend on the scattering phase $\phi_s$, while the quasi-periodic oscillating behavior is controlled by the coupling energy $\Gamma$.
The above behavior is peculiar for a gapped spectrum of the quasi-particles, while for large enough transverse dimension $W$ the maxima position does not depend on $\phi_s$, as in the case of massless Dirac particles. A similar behavior is also observed in the $J_c$ \textit{vs} $d$ curves, where an oscillating dependence on the constriction length is present.
The effect of $\phi_s$ on the transport properties is also evident in the SBC configuration. In particular, the $J_s$ vs $eV_g$ curves present an oscillating pattern reminiscent of the quantum interference, while the behavior of $J_s$ \textit{vs} $d$ curves is qualitatively similar to the one observed for the charge current in CBC. However, the spin current is strongly dependent on the spin-flipping tunneling value at the etching points $x_{1,2}$. In particular, the increasing of $\gamma_{sf}$ above a certain threshold value induces a change of sign of $J_s$, the latter being important to characterize the interface properties of the device and for spintronics purposes.\\
Concerning the gate configuration (ii), in the CBC we observe  that the $J_c$ \textit{vs} $eV_g$ curves do not depend on $eV_g$. This behavior can be understood observing that the Fabry-P\'{e}rot phase $\phi_{FP}$ responsible for the gate modulation of $J_c$ is zero. On the other hand, the $J_c$ \textit{vs} $d$ curves present a behavior similar to the one observed in the gate configuration (i).\\
In SBC, the $J_s$ \textit{vs} $eV_g$ curves present an oscillating behavior induced by an Aharonov-Bohm-like phase $\phi_{AB}\neq 0$, which is strongly affected by the tunneling amplitudes $\gamma_{sp}$ and $\gamma_{sf}$. In particular for high values of $\gamma_{sf}$, a full electrical switching of the spin current can be obtained by tuning $eV_g$ in the interval $[-3.9,-3.2]$ (see Fig.\ref{fig:fig11}). Indeed, using the external gates, it is possible to tune the spin current from $J_s=-2$ ($eV_g=-3.9$) up to $J_s=0.8$ ($eV_g=-3.2$), being this effect relevant for spintronics.
Moreover, the $J_s$ \textit{vs} $d$ curves present a quantum beating-like structure caused by the presence of two (non-commensurate) periods of oscillation, namely  $\tau_d^{(1/2)}$. Finally, there exist special values of $d$ for which the spin current is independent from $\gamma_{sf}$ (see Fig.\ref{fig:fig12}).\\
In conclusion, we have shown that the electrical switching behavior offers an efficient and robust mechanism to manipulate topological edge states transport
exploiting a four terminal set-up relevant for spintronics applications. Furthermore, going beyond the technological implications, our analysis can be used to unveil the effects of a non-vanishing effective mass on the transport properties of confined Dirac Fermions.

\appendix
\section{Equation of motion and model assumptions}
\label{app: appA}
\begin{figure}[t]
\centering
\vspace{0.5cm}
\includegraphics[clip,scale=0.65]{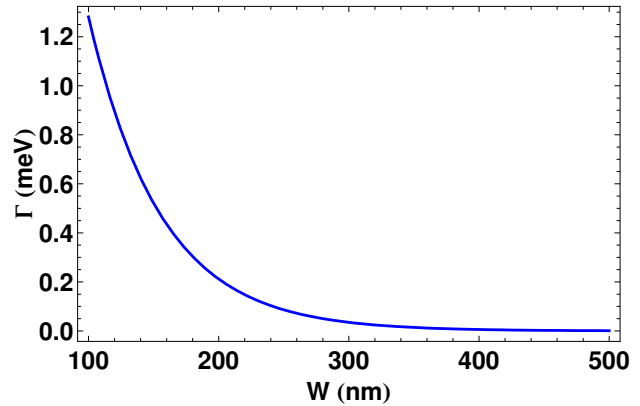}\\
\caption{(Color online) Exponential dependence of $\Gamma$ on W as deduced from Ref.[\onlinecite{niu_prl_2008}], the interpolation formula being $\Gamma\approx \Gamma_0 \exp(-\lambda W)$, with $\Gamma_0=7.75$ meV and $\lambda=0.018$ nm$^{-1}$.}
\label{fig:fig13}
\end{figure}

Within the constriction ($x_1<x<x_2$), the equation of motion of $\Psi(x)=(\psi_{R\uparrow},\psi_{R\downarrow},\psi_{L\uparrow},\psi_{L\downarrow})^t$ is governed by the Hamiltonian $H=H_0+H_{c}$ where for generality, we assume that apart from the term (\ref{eq:coupling}) a term with spin-flipping coupling of the form (\ref{eq:spin-flip}) with constant amplitude $\Gamma_f$ is also present:
\begin{equation}
i\hbar\partial_t \Psi(x)=\left[
                           \begin{array}{cccc}
                             \mathcal{D}_{+} & \Gamma_f & \Gamma & 0 \\
                             \Gamma_f & \mathcal{D}_{+} & 0 & \Gamma \\
                             \Gamma & 0 & \mathcal{D}_{-} & -\Gamma_f \\
                             0 & \Gamma & -\Gamma_f & \mathcal{D}_{-} \\
                           \end{array}
                         \right]
\Psi(x),
\end{equation}
with $\mathcal{D}_{\pm}=\mp i\hbar v_F \partial_x$. Considering constrictions whose transverse dimension $W$ allows edge states coupling, the functional form of the coupling\cite{niu_prl_2008} is $\Gamma\approx \Gamma_0 \exp(-\lambda W)$ (see Fig.\ref{fig:fig13}), while the flipping term $\Gamma_{f}$ can be neglected for not too tight and not too long constrictions. As also evident from Ref.[\onlinecite{richter_prl_2011}], the spin precession induced by a very weak value of the spin-orbit interaction assumes a relevant role in the transport only for very long constrictions characterized by  $L \approx 900-1900$ nm which are quite long to maintain  coherent transport. This comes from the fact that a slow spin precession needs a very long time (and thus a long distance) to completely change the spin polarization of a particle traveling along the system. In particular, a spin precession described by an angular frequency $\omega_{SO}$ requires a time $\tau=\pi/\omega_{SO}$ to completely flip the electron spin, while the dwell time $\tau_{dw}$ required to cover the constriction length $L$ is given by $L/v_F$. The above condition implies that the shortest constriction length to observe a complete spin flipping fulfills the relation $L=\pi v_F/\omega_{SO}$. As a consequence, weak values of the spin orbit coupling (and thus small values of $\omega_{SO}$) require longer constrictions. Motivated by these arguments, in this work we assume $L<1 \ \mu$m and set $\Gamma_f=0$ along the constriction, which is equivalent to neglect the spin-orbit coupling, while we consider spin-flipping tunneling $\gamma_{sf}$ at the etching points $x=x_{1/2}$.

%----------------------------------------------------------fig13
%----------------------------------------------------------

\section{Transition probabilities}
\label{app: appB}
\begin{figure}[!h]
\centering
\includegraphics[clip,scale=0.325]{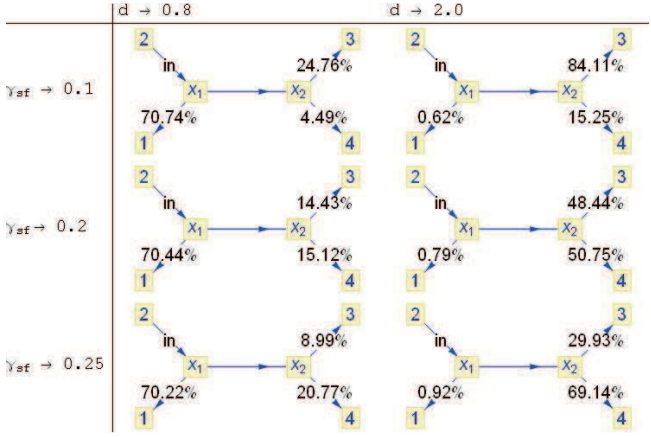}\\
\caption{(Color online) Transition Probabilities $P_{2 \rightarrow f}=|S_{f2}|^2$ from the injection lead 2 ($in$) to the arrival lead $f\in\{1,3,4\}$. The parameters are fixed as done in Fig.\ref{fig:fig6} ($eV_g=-2.45$, $\Gamma=2$, $\gamma_{sp}=0.1$), while the values of $d$ and $\gamma_{sf}$ are specified by the column/row labels. Notice the resonant transmission regime, characterized by $P_{2 \rightarrow 1 }<1\%$, obtained by fixing the constriction length $d=2$.}
\label{fig:fig14}
\end{figure}

In this Appendix we show how topological edge states can be selectively switched in the nanoconstriction. In Fig.\ref{fig:fig14} are shown the transition probabilities for an incoming spin-polarized state at the upper edge (terminal 2) to be reflected back to the lower edge (terminal 1), transmitted through the constriction in the same spin and edge state (terminal 3) or transmitted by swapping the edge and simultaneously flipping the spin (terminal 4). By using the parameters $eV_g=-2.45$, $\Gamma=2$, $\gamma_{sp}=0.1$ (as given in Fig.\ref{fig:fig6}), and fixing the electrochemical potential at the bottom of the conduction band, one observes that transmission from a state up to a state down in terminal 4 is activated for longer nanoconstriction channels and higher values of the local spin flipping tunneling at the point contacts (see right down panel), while for lower values of $\gamma_{sf}$ transmission along the same edge and spin state is favored for longer channels (see panel right up).
%----------------------------------------------------------fig14
%----------------------------------------------------------
For intermediate values of $\gamma_{sf}$ both reflection to the lower edge in the same spin state and transmission along the channel in terminal 3 and 4 takes place, the latter having almost equal probabilities (see middle panels). Since the spin-flipping tunneling is controlled by the local modification of the spin-orbit interaction by geometrical etching, our device can be used as spin transistor with high fidelity. Let us note that differently from Ref.[\onlinecite{richter_prl_2011}], here spin precession along the channel does not take place and is not relevant for this setup.

%================================================================ack
\section*{Acknowledgements}
We thank A. Braggio, G. Dolcetto, and N. Magnoli for useful discussions. The support of CNR STM 2010 program, EU-FP7 via Grant No. ITN-2008- 234970 NANOCTM and CNR-SPIN via Seed Project PGESE001 is acknowledged.
%=======================================================================================bibliografia
\bibliographystyle{prsty}

\begin{thebibliography}{99}
\bibitem{note1} Where not specified the integration domain is given by $[-\infty,+\infty]$, i.e. $\int dx \equiv \int_{-\infty}^{\infty} dx$.
\end{thebibliography}

%===============================================================================================bib
\end{document}